# Real-time observation of a coherent lattice transformation into a high-symmetry phase

Samuel W. Teitelbaum[1,†,*], Taeho Shin[2,†], Johanna W. Wolfson[1], Yu-Hsiang Cheng[3], Ilana J. Porter[1], Maria Kandyla[1,4] and Keith A. Nelson[1,*]

[1] Department of Chemistry, Massachusetts Institute of Technology, 77 Massachusetts Ave., Cambridge MA, 02139, USA

[2] Department of Chemistry, Chonbuk National University, Jeonju 54896, Republic of Korea

[3] Department of Electrical Engineering and Computer Science, Massachusetts Institute of Technology, 77 Massachusetts Ave., Cambridge MA, 02139, USA

[4]Theoretical and Physical Chemistry Institute, National Hellenic Research Foundation, 48 Vasileos Constantinou Ave., Athens 11635, Greece

*Correspondence to: kanelson@mit.edu, samuelt@mit.edu

†These authors contributed equally to this work

**Excursions far from their equilibrium structures can bring crystalline solids through collective transformations including transitions into new phases that may be transient or long-lived. Direct spectroscopic observation of far-from-equilibrium rearrangements provides fundamental mechanistic insight into chemical and structural transformations, and a potential route to practical applications, including ultrafast optical control over material structure and properties. However, in many cases photoinduced transitions are irreversible or only slowly reversible, or the light fluence required exceeds material damage thresholds. This precludes conventional ultrafast spectroscopy in which optical excitation and probe pulses irradiate the sample many times, each measurement providing information about the sample response at just one probe delay time following excitation, with each measurement at a high repetition rate and with the sample fully recovering its initial state in between measurements. Using a single-shot, real-time measurement method, we were able to observe the photoinduced phase transition from the semimetallic, low-symmetry phase of crystalline bismuth into a high-symmetry phase whose existence at high electronic excitation densities was predicted based on earlier measurements at moderate excitation densities below the damage threshold. Our observations indicate that coherent lattice vibrational motion launched upon photoexcitation with an incident fluence above 10 mJ/cm$^2$ in bulk bismuth brings the lattice structure directly into the high-symmetry configuration for tens of picoseconds, after which carrier relaxation and diffusion restore the equilibrium lattice configuration.**

# Introduction

Crystals with symmetry-lowering Peierls distortions[1–5] present particularly attractive targets for time-resolved measurements of collective transitions to new structural phases. Like their molecular counterparts with Jahn-Teller distortions, Peierls-distorted systems arise due to coupling between electrons in partially filled states of nearby energy and vibrational modes along whose coordinates the distortions occur. The energy cost of the vibrational excursion is more than compensated for by the associated splitting of degenerate or near-degenerate electronic states, with electrons populating the lower-energy states. Optical excitation removes electrons from those states, reducing the energy payback for vibrational distortion. Under sufficiently intense illumination, the distortion might be removed completely. The semimetal is a prototypical Peierls system[6,7]; where its rhombohedral crystal lattice is related to a simple cubic structure (observed in antimony at elevated pressures as the stable, metallic phase[8,9]) through motion of atoms away from the high-symmetry locations along the unit cell diagonals, corresponding to the $A_{1g}$ optic phonon mode of the rhombohedral phase. This collective distortion opens a bandgap to produce the semimetallic phase. Previous ultrafast spectroscopic studies on bismuth have shown time-dependent oscillations of the $A_{1g}$ phonon mode, with significant softening of the phonon frequency as the excitation density was increased up to 1.5% of the valence electrons[8]. Density functional theory (DFT) calculations by Murray et al.[4] predict that if the electronic excitation is strong enough (>2% of valence electrons excited), it should be possible to drive the bismuth atom past the center of the unit cell and reach a transient high-symmetry structure, shown in Fig. 1a. This photoinduced phase transition, which has not yet been observed, is predicted to be nonthermal, and the high-symmetry phase cannot be reached by raising the temperature. The lattice change occurs due to exceptionally strong electron-phonon coupling, as removal of valence electrons upon optical absorption changes the lattice potential preceding any loss of electron energy to the lattice. In contrast, in aluminum and many other metals, carrier excitation has minimal effect on the lattice potential energy, and phase transitions such as lattice melting proceed through the transfer of energy from hot carriers to the initially cold lattice over a few picoseconds[11]. The symmetry-lowering removal of the Peierls distortion by electronic excitation has been observed in other Peierls-like materials, such as $VO_2$[12], blue bronze[5], charge-ordered manganites[13–16], organic crystals[1,3,17], and rare-earth chalcogenides[18,19]. Despite bismuth's status as a canonical Peierls system, sample damage at high excitation densities has prevented observation of electronic lifting of the Peierls distortion analogous to what has been observed in similar systems. Confirmation that the Peierls distortion in bismuth can be lifted with electronic excitation would provide a benchmark for electronic lifting of a Peierls distortion in a relatively well-understood system with minimal electron correlation, where the observed dynamics are purely structural.

Fritz et al. performed optical pump x-ray probe measurements in bismuth, showing bond softening and approach toward the symmetric phase[10]. However, sample damage prevented exploration at excitation densities beyond where 1.3% of the valence electrons are excited. Sciaini et al. probed photoexcited bismuth with femtosecond electron diffraction and showed evidence of nonthermal melting faster than 1 ps after high-fluence excitation above the damage threshold[20], and Sokolowski-Tinten et al. reported X-ray diffraction measurements that showed a loss of long-range order after several picoseconds [21]. The results to date leave open the question of whether a transient high-symmetry crystalline state in bismuth may be reached upon photoexcitation. In this "hidden" phase, the unit cell would be halved in size, with only one atom per unit cell and no optic phonon modes. Therefore, the absence of coherent optical phonon oscillations in a time-resolved pump-probe trace with high pump fluence could indicate a higher symmetry crystalline phase with no long-range order between bismuth dimers. However, the Raman-active $A_{1g}$ optic phonon mode is also absent in liquid and amorphous solid systems. Lowering the initial sample temperature will preclude thermal melting at high laser fluence. The time-dependent reflectivity can then monitor the time-dependent evolution of the lattice distortion closely, but such data are difficult to collect if each high-fluence excitation laser shot is followed by a single probe pulse that records the transient optical properties at a single delay time. In that case, permanent damage of each irradiated sample region allows data to be recorded only at a few time delays, with limited signal/noise ratio because extensive signal averaging is impossible and the time dependence must be determined by combining measurements at different delays from different sample regions. Here we report results of measurements that circumvent these limitations by recording the sample response at many delay times after each excitation pulse[22], providing an experimental probe of a "hidden" solid phase and measurement of the dynamics of its formation and relaxation.

## Methods

The bismuth samples were polycrystalline thin films sputtered on sapphire and glass substrates, with film thicknesses of 20 nm and 275 nm (essentially bulk). The substrate was heated to 170 °C during deposition to ensure crystalline deposition. Substrates were sonicated in acetone to remove any residual surface deposits, then plasma cleaned in the sputter coater. Powder diffraction confirmed the samples were crystalline and slightly textured with the surface preferentially oriented along the (1 1 1) direction.

Single-shot transient reflection pump-probe measurements were conducted using a method demonstrated earlier[22] in which a 20 x 20 grid of 400 probe pulses is focused onto the same region of the sample as an 800 nm excitation pulse, with the probe pulse time delays incremented by 23 fs. The reflected probe pulses are projected onto a CCD camera where they again form a 20 x 20 grid, allowing the data from each of the 400 time delays to be read out. The method is summarized in the supplementary material, and representative images and traces are shown in Fig. 1b and Fig. 1c, respectively. To ensure that melting could not occur before formation of the symmetric phase, the sample was cooled to 80 K, well below bismuth's melting temperature of 544 K. Depending on the pump fluence, 20-100 shots were recorded from each sample region before permanent damage occurred, as assessed by changes in static reflectivity and in pump-probe time-dependent signals that were recorded at low fluence, and confirmed after the experiment by checking for visible damage on the sample surface with an optical microscope. The pump spot size was approximately 1 mm in diameter on the sample, and was determined by placing a camera at the sample position.

## Results

**Single-shot Spectroscopy**

Single-shot data from the bulk (275 nm film) and thin film (20 nm film) samples are shown in Fig. 2. Reflectivity traces for the bulk sample are shown in Figs. 2a and b, while similar traces for the 20 nm thin film are shown in Fig. S6 at the Supplementary Information. In bulk bismuth, upon increasing the excitation fluence, the lifetime of the coherent phonon mode drops dramatically, and the frequency drops from its equilibrium value of 2.95 THz down to a minimum measurable value of 1.8 THz (Fig 2c, closed circles). Above 10 mJ/cm$^2$, the coherent phonon oscillations persist for less than a half cycle (Fig. 2b), indicating that the coherent lattice vibrational motion initiated upon photoexcitation through electron-phonon coupling propels the atoms directly into their positions in the high-symmetry phase. Above this fluence, the initial potential energy along the vibrational coordinate in the excited state is higher than the barrier in the center of the unit cell, as shown in Fig. 1a. The initial motion of the bismuth atoms toward (and past) the center of the unit cell, followed by rapid dephasing, randomizes the Bi atom positions in their local potential minima about the unit cell centers, thereby forming the high-symmetry phase.

We see similar results in bond softening and the eventual disappearance of coherent oscillations at all temperatures between 300 K and 80 K in both the 20 nm and 275 nm thin films. In the 20-nm film, lower pump fluences are required for similar reductions in phonon frequency and for the disappearance of the phonon oscillations (Fig. 2c, open circles). As has been discussed earlier[23], confinement of excited carriers in the thin film preserves a higher carrier density than in the thicker sample, in which fast carrier diffusion quickly reduces the photoexcited electron density. This accounts for more pronounced thin film responses at lower pump fluences, as has been noted in several studies[10,24]. Quantum confinement and back-reflections from the substrate in the 20-nm thin film reverse the sign of the reflectivity change and reduce the amplitudes of the reflectivity oscillations compared to the bulk sample[23], but do not affect the excitation density-dependence of the lattice potential. The fluence-dependent trend in bond softening does not change significantly with sample temperature between 300 K and 80 K (Fig. 2d). Small differences may be due to thermal contraction and a slightly higher carrier diffusivity at low temperature.

To measure coherent phonon frequencies in double-pump and single-pump pump-probe experiments, single-shot traces were fit to a decaying background signal of the form

$$\left(\frac{\Delta R}{R}\right)_{bgd} = A_1 e^{-t/\tau_{el}} + A_2 \qquad 1.$$

Where $\tau_{el}$ is the decay time of the slowly varying background, and $A_1$ and $A_2$ are the amplitudes of the decaying and quasi-DC components. To obtain the phonon frequency, the trace was subtracted from the fitted background and fit to a decaying sine function of the form

$$\left(\frac{\Delta R}{R}\right)_{osc} = A \cos(2\pi f t) e^{-\gamma t} \qquad 2.$$

where $f$ is the coherent $A_{1g}$ phonon frequency, $A$ is the amplitude of the phonon oscillation, and $\gamma$ is the phonon damping rate. The chirp in the $A_{1g}$ mode oscillation observed in previous studies[4,25] is not necessary for fitting traces under very high fluence excitation and most two-pulse data after high fluence excitation because the phonon lifetime is sufficiently short that no change in the phonon frequency is measurable during the decay time. Fits (black lines) are shown superimposed on the raw data (color) for each single-shot trace as a function of fluence in Fig. 2a and 2b. The fitted frequency $f$ and damping rate $\gamma$ as a function of fluence are shown in Fig. 2c and 2d, respectively. The fitting breaks down at high fluence when the coherent oscillation is seen for less than a half cycle, very close to and above the onset of the symmetric phase.

As discussed below, thermal modeling shows that an excitation fluence of 20 mJ/cm$^2$ or below is not sufficient to bring the surface of the 275-nm sample from 80 K to the melting point, indicating that the loss of phonon signal is due to transition into the high-symmetry crystalline phase rather than melting. To probe the structural evolution of the bismuth crystal after strong photoexcitation, we irradiated the sample with a second, weak excitation pulse at various time delays following the first strong pump pulse, and we measured the time-dependent signal following the second pulse. This allowed us to observe the recovery of phonon oscillations induced by a weak pump pulse, indicating if the crystal had returned to the low-symmetry phase by the time that the second pulse arrived[12]. Figure 3a shows results from the 275-nm film at 80 K for a first pump fluence of 10.9 mJ/cm$^2$ and a second pump fluence of 1 mJ/cm$^2$. After a delay of 3 ps between the first and the second pump pulse, there is no indication of coherent oscillations induced by the second pulse. After a 20-ps delay between the first and second pump pulses, the coherent oscillations are clearly present. For the initial pump fluence of 10.9 mJ/cm$^2$, the phonon oscillations return within 10 ps. For lower initial pump fluences the oscillation recovers even sooner. The oscillation frequency as a function of the delay of the second pump pulse with respect to the first pump pulse, for two fluence values of the first pump pulse, is shown in Fig. 3b. The low first pump pulse fluence of 7 mJ/cm$^2$ is below the transition threshold to the symmetric phase and the higher pump pulse fluence of 10 mJ/cm$^2$ is above the transition threshold. The picosecond time scale for restoration of the low-symmetry crystalline state is consistent with the kinetics of electron-hole recombination and diffusion, but is clearly inconsistent with recrystallization from a liquid state either through nucleation or growth from nearby crystalline regions, which would be on the order of nanoseconds or more. This indicates that the disappearance of the Peierls distortion as measured by the absence of an $A_{1g}$ coherent phonon is not due to the formation of a liquid phase through either thermal or non-thermal melting. Results of a two-pump experiment on the same film with an initial sample temperature of 300 K are shown in Fig. 3d, showing a similar recovery timescale of the coherent phonon frequency. The symmetric phase is reached for the trace with an initial pump fluence of 10.9 mJ/cm$^2$.

Furthermore, the single-shot instrument permits us to measure the time-dependent reflectivity after the initial, strong excitation pulse for up to 1 ns following photoexcitation. The final time-dependent reflectivity change can be used as a secondary check of the final lattice temperature after a strong photoexcitation pulse, using the known thermoreflectance of bismuth. The time-dependent reflectivities at 800 nm at various pump fluences are shown in Fig. 3c. For a probe wavelength of 800 nm, the thermoreflectance parameter in bismuth is $dR/dT = 9\times10^{-4}$ K$^{-1}$ [26]. Therefore, the 4% change in reflectivity after excitation with a 30 mJ/cm$^2$ pulse corresponds to a 500 K temperature jump. This is in good agreement with our thermal model at later time delays ( >60 ps) as discussed in the next section.

## Modeling and Simulations

We can calculate the expected carrier density within the optical penetration depth based on the excitation fluence and the carrier diffusion rate. Then, we can use the potential energy surface calculated by Murray et al.[4] to relate the carrier density at the irradiated sample surface to the phonon frequency. Based on our calculations for carrier diffusion and the measured phonon frequency, the maximum photoexcited carrier density reached at the surface in the experiments presented here is greater than 2% of the valence electrons excited. This is above the calculated threshold for the structural phase transition in bulk Bi. The photoexcitation of 2% valence electrons threshold is reached with a pump fluence of 13 mJ/cm$^2$.

Using an extended two-temperature model[27], we simulated carrier diffusion and thermalization away from the irradiated sample surface and thermalization with the lattice, the results of which are shown in Fig. 4. The parameters used for the extended two-temperature model agree well with previous results at moderate excitation density[10,23,24,28]. We found that the maximum lattice temperature $T_{max}$ is reached 15 ps after excitation, and 90% of the lattice temperature jump is complete within 5 ps. This temperature rise is consistent with similar results using a conventional two-temperature model[28,29] which does not take into account independent carrier density and temperature. The lattice would reach the melting point $T_m$ with enough excess heat to surpass the heat of fusion in the heated volume only with a far higher fluence than that required to reach the symmetric phase at room temperature. At 80 K, the surface temperature reaches the melting point at a comparable excitation fluence to the onset of the high-symmetry phase transition, but is never heated past the heat of fusion required for melting to occur over the probe penetration depth (15 nm). In order for a liquid phase to form, the bismuth film must be superheated well past the melting point order to overcome the activation energy required to form a solid-liquid interface[30]. Based on the thermal model, $T_{max}/T_m$ is 1.1, which results in melting times well over one second using homogeneous nucleation theory[30]. A small liquid layer could form near the surface in tens of ps, but cannot account for the complete loss of coherent phonon signal at high excitation densities.

Based on the carrier density and lattice temperature calculated using our transport and thermalization models, we can predict the $A_{1g}$ phonon frequency after excitation, and compare with the experimentally determined $A_{1g}$ phonon frequency. The comparison is shown between the points and curves in Fig. 3b, which agree to within the uncertainties of our fitting. As the photoexcited carriers recombine, the lattice distortion and coherent phonon frequency recover toward their equilibrium values. The persistence of the phonon frequency redshift at long times (>60 ps) is due to heating of the bismuth film[25]. The predicted coherent phonon frequencies based on the lattice temperature (Fig. 4a) and carrier density (Fig. 4b) simulated by an extended two-temperature model agree well with the measured phonon frequencies.

Similarly, we can use the final reflectivity after the strong excitation pulse as a secondary thermometer at long delays, given the known thermoreflectance of bismuth. The time-dependent reflectivity is shown in Fig. 3c. In comparison with our model, the reflectivity agrees with our model to within 20% at delays longer than 80 ps. Given the accuracy of the thermoreflectance parameter, this is effectively within the margin of error of our model, and demonstrates that our thermal model correctly predicts the lattice temperature at late times. At early times, when hot carriers persist and have not thermalized with the lattice, we would not expect the lattice temperature alone to describe the transient reflectivity change of the sample. Nevertheless, the agreement at late times indicates that the thermal model is able to estimate the overall lattice heating 80 ps after the excitation pulse. The expected time for minimum reflectivity based on thermoreflectance is 20 ps, earlier than what is observed experimentally, where the minimum reflectivity is around 60 ps. This is likely due to competition between hot carriers, which result in a rise in reflectivity, and heating, which results in a drop in reflectivity, and not necessarily due to an underestimate of the heating time.

Finally, we simulated the process by which the bismuth atoms are driven across the center of the unit cell. This barrier crossing, when combined with strong damping, destroys the Peierls distortion and forms the symmetric phase. To simulate this process, we solve the classical equations of motion numerically for a particle in a slowly varying potential. At every point in time, we use the appropriate lattice potential from the time-dependent carrier density derived from the extended two-temperature model. This, along with an empirical damping rate based on the incident fluence (Fig. 2d), allows us to solve the equations of motion for the coherent part of the $A_{1g}$ mode trajectory $x(t)$. This trajectory can then be converted to a reflectivity change by

$$\left(\frac{\Delta R}{R}\right) \propto x(t)^2 - x_0^2 \qquad\qquad 3.$$

where $x_0$ is the initial position of the Bi atom along the c-axis direction, and $x(t)$ is the time-dependent displacement, where $x=0$ is the center of the unit cell. These trajectories are then fit to a decaying sine function plus a slowly varying background using the same procedure as in the pump-probe trace. Several simulated trajectories for incident fluences used in the experiments shown in Fig. 2a are shown in Fig. 4c. A comparison of the fitted parameters for the simulated trajectories and data is shown in Fig. 4d and Fig. 4e. Note that the fits break down when the barrier is crossed, as highly anharmonic behavior, strong coupling to other modes[6], and a loss of Raman activity of the $A_{1g}$ mode[5,12] are expected in this fluence regime.

## Discussion

Having presented the experimental evidence for the appearance of a high-symmetry crystalline phase of bismuth under strong photoexcitation and thermal modeling that estimates the extent of electronic and thermal excitation at the sample surface, we now return to our picture of the formation of the high-symmetry phase. The key phenomena and timescales are the following, described approximately by the two-temperature model that has been applied to bismuth[19]. Within the excitation pulse duration, the excited electrons equilibrate with each other at the 1.5-eV photon energy, establishing an excited carrier temperature of about 10,000 K. Carriers diffuse out of the 15-nm optical penetration depth of the crystal while electron-phonon coupling leads to heating of the lattice, which reaches a maximum temperature in the irradiated region after approximately 15 ps, consistent with optical and electronic spectroscopy measurements[10,24]. Thermal diffusion on nanosecond time scales will finally return the crystal surface to the initial temperature. Even well below the threshold for the photoinduced phase transition, the $A_{1g}$ phonon frequency will reflect these kinetics. The initial high carrier density leads to pronounced phonon softening, the extent of which decreases as the excited electrons relax and diffuse out of the optical penetration depth. The elevated lattice temperature induces a more moderate phonon frequency reduction due to lattice anharmonicity. With moderate pump fluences, such as 6.9 mJ/cm$^2$, the phonon frequency recovers to its equilibrium value of 2.9 THz after less than 100 ps, as measured with a second weak pump pulse at long delay times (see Fig. 3b). At higher pump fluences, the phonon frequency does not fully recover even after the longest second-pump delays of 800 ps shown in Fig. 3b, indicating an elevated lattice temperature that persists for nanoseconds or longer.

In the short time (>150 fs) it takes to form the high-symmetry phase, the unit cell does not have time to change volume, so the primitive lattice likely remains rhombohedral. Within tens of picoseconds following excitation above the threshold for the phase transition, carrier diffusion and thermalization partially restore the lattice potential, which restores lattice ordering along the $A_{1g}$ mode coordinate, resulting in reappearance of the $A_{1g}$ coherent phonon response after a weak second excitation pulse.

We excited bulk and thin film bismuth with high fluences, and through the use of a real-time measurement method we observed an ultrafast phase transition into a high-symmetry crystalline phase. Two-pulse measurements show that the symmetric phase persists for several picoseconds before the lattice distortion recovers. Our results show that the predicted high-symmetry crystalline phase, which is inaccessible thermally, exists, and that it is reached through coherent collective motion of the photoexcited lattice. Direct single-shot observation of coherent motion all the way to far-from-equilibrium structures despite the photoexcited sample never returning to its pristine state has enabled comparison to first-principles theoretical calculations and opens up possibilities for coherent control that could be extended to multi-dimensional control in materials with complex phase diagrams.

**References and Notes:**

1. Chollet, M. *et al.* Gigantic photoresponse in 1/4-filled-band organic salt (EDO-TTF)2PF6. *Science* **307,** 86–9 (2005).

2. Matsubara, M. *et al.* Photoinduced switching between charge and orbital ordered insulator and ferromagnetic metal in perovskite manganites. *Phys. Rev. B - Condens. Matter Mater. Phys.* **77,** 1–5 (2008).

3. Iwai *et al.* Ultrafast optical switching from an ionic to a neutral state in tetrathiafulvalene-p-chloranil (TTF-CA) observed in femtosecond reflection spectroscopy. *Phys. Rev. Lett.* **88,** 57402 (2002).

4. Murray, É., Fritz, D., Wahlstrand, J., Fahy, S. & Reis, D. Effect of lattice anharmonicity on high-amplitude phonon dynamics in photoexcited bismuth. *Phys. Rev. B* **72,** 60301 (2005).

5. Huber, T. *et al.* Coherent structural dynamics of a prototypical charge-density-wave-to-metal transition. *Phys. Rev. Lett.* **113,** 1–5 (2014).

6. Zijlstra, E. S., Tatarinova, L. L. & Garcia, M. E. Laser-induced phonon-phonon interactions in bismuth. *Phys. Rev. B* **74,** 220301 (2006).

7. Jones, H. Applications of the Bloch Theory to the Study of Alloys and of the Properties of Bismuth. *Proc. R. Soc. A Math. Phys. Eng. Sci.* **147,** 396–417 (1934).

8. Beister, H. J., Strössner, K. & Syassen, K. Rhombohedral to simple-cubic phase transition in arsenic under pressure. *Phys. Rev. B* **41,** 5535–5543 (1990).

9. Kabalkina, S. S., Kolobyanina, T. N. & Vereshchagin, L. F. Investigation of the Crystal Structure of the Antimony and Bismuth High Pressure Phases. *Sov. J. Exp. Theor. Phys.* **31,** 259 (1970).

10. Fritz, D. M. *et al.* Ultrafast bond softening in bismuth: mapping a solid's interatomic potential with X-rays. *Science* **315,** 633–6 (2007).

11. Siwick, B. J., Dwyer, J. R., Jordan, R. E. & Miller, R. J. D. An atomic-level view of melting using femtosecond electron diffraction. *Science* **302,** 1382–5 (2003).

12. Wall, S. *et al.* Ultrafast changes in lattice symmetry probed by coherent phonons. (2012). doi:10.1038/ncomms1719

13. Zhang, J. *et al.* Cooperative photoinduced metastable phase control in strained manganite films. *Nat. Mater.* **15,** 956–960 (2016).

14. Beaud, P. *et al.* A time-dependent order parameter for ultrafast photoinduced phase transitions. *Nat. Mater.* **13,** 923–7 (2014).

15. Beaud, P. *et al.* Ultrafast structural phase transition driven by photoinduced melting of charge and orbital order. *Phys. Rev. Lett.* **103,** 155702 (2009).

16. Matsubara, M. *et al.* Photoinduced switching between charge and orbital ordered insulator and ferromagnetic metal in perovskite manganites. *Phys. Rev. B* **77,** 94410 (2008).

17. Koshihara, S., Tokura, Y., Takeda, K. & Koda, T. Reversible photoinduced phase transitions in single crystals of polydiacetylenes. *Phys. Rev. Lett.* **68,** 1148–1151 (1992).

18. Schmitt, F. *et al.* Transient electronic structure and melting of a charge density wave in TbTe3. *Science* **321,** 1649–52 (2008).

19. Stojchevska, L. *et al.* Ultrafast Switching to a Stable Hidden Quantum State in an Electronic Crystal. *Science (80-. ).* **344,** 177–180 (2014).

20. Sciaini, G. *et al.* Electronic acceleration of atomic motions and disordering in bismuth. *Nature* **458,** 56–9 (2009).

21. Sokolowski-Tinten, K. *et al.* Femtosecond X-ray measurement of coherent lattice vibrations near the Lindemann stability limit. *Nature* **422,** 287–289 (2003).

22. Shin, T., Wolfson, J. W., Teitelbaum, S., Kandyla, M. & Nelson, K. A. Dual Echelon Femtosecond Single-Shot Spectroscopy. *Rev. Sci. Instrum.* **85,** (2014).

23. Shin, T., Wolfson, J. W., Teitelbaum, S. W., Kandyla, M. & Nelson, K. A. Carrier confinement and bond softening in photoexcited bismuth films. *Phys. Rev. B - Condens. Matter Mater. Phys.* **92,** 1–5 (2015).

24. Liebig, C., Wang, Y. & Xu, X. Controlling phase change through ultrafast excitation of coherent phonons. *Opt. Express* **18,** 633–636 (2010).

25. Garl, T. *et al.* Birth and decay of coherent optical phonons in femtosecond-laser-excited bismuth. *Phys. Rev. B* **78,** 134302 (2008).

26. Wu, A. Q. & Xu, X. Coupling of ultrafast laser energy to coherent phonons in bismuth. *Appl. Phys. Lett.* **90,** 23–25 (2007).

27. Shin, T., Teitelbaum, S. W., Wolfson, J., Kandyla, M. & Nelson, K. A. Extended two-temperature model for ultrafast thermal response of band gap materials upon impulsive optical excitation. *J. Chem. Phys.* **143,** 194705 (2015).
28. Arnaud, B. & Giret, Y. Electron Cooling and Debye-Waller Effect in Photoexcited Bismuth. *Phys. Rev. Lett.* **110,** 16405 (2013).

29. Sokolowski-Tinten, K. *et al.* Thickness-dependent electron–lattice equilibration in laser-excited thin bismuth films. *New J. Phys.* **17,** 113047 (2015).

30. Rethfeld, B., Sokolowski-Tinten, K., von der Linde, D. & Anisimov, S. Ultrafast thermal melting of laser-excited solids by homogeneous nucleation. *Phys. Rev. B* **65,** 92103 (2002).

31. Sheu, Y. M., Chien, Y. J., Uher, C., Fahy, S. & Reis, D. A. Free-carrier relaxation and lattice heating in photoexcited bismuth. *Phys. Rev. B* **87,** 75429 (2013).

32. Sheu, Y. M. *et al.* Kapitza conductance of Bi/sapphire interface studied by depth- and time-resolved X-ray diffraction. *Solid State Commun.* **151,** 826–829 (2011).
**Acknowledgments:** The authors would like to thank A. Maznev for helpful discussions regarding thermal modeling. This work was also supported by Office of Naval Research Grants No. N00014-12-1-0530 and N00014-16-1-2090 and the National Science Foundation Grant No. CHE-1111557. This work was performed in part at the Center for Nanoscale Systems (CNS), a member of the National Nanotechnology Infrastructure Network (NNIN), which is supported by the National Science Foundation under NSF award no. ECS-0335765. CNS is part of Harvard University. T.S. acknowledges the support from the National Research Foundation (NRF-2016R1C1B2010444).

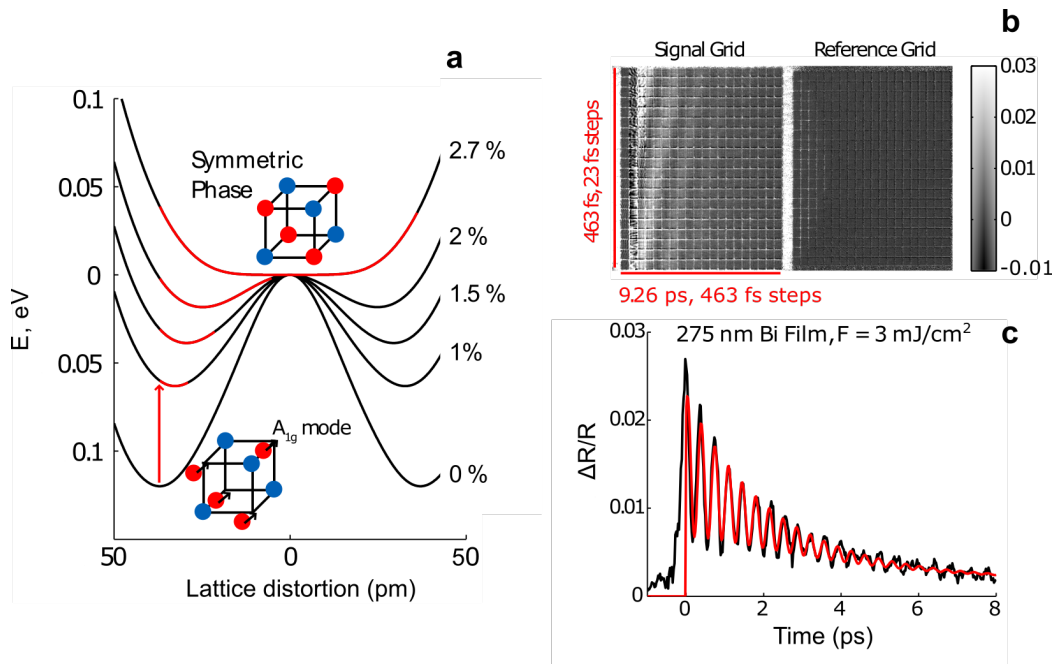

**Fig. 1**. Dual-echelon single-shot spectroscopy of bismuth. **a** Potential energy surface of bismuth along the $A_{1g}$ mode, at various levels of electronic excitation, offset for clarity. The range of lattice motion upon impulsive excitation from the ground state is shown in red. At excitation densities above 2% of the valence electrons, the kinetic energy imparted to the $A_{1g}$ mode is enough to drive the central atom across the center of the unit cell, forming a symmetric phase. The distorted and symmetric crystal structures are shown in part **a**. The lattice distortion is exaggerated for clarity. **b** Differential image of the echelon grid on a 275 nm thick bismuth film, excited with a fluence of 3 mJ/cm$^2$ and a probe wavelength of 800 nm, on a scale of -1 to 3% $\Delta R/R$, showing coherent phonon oscillations on the signal (left) grid. This image was averaged over 50 laser shots. **c** Unfolded single-shot pump-probe trace extracted from the image in A, showing coherent phonon oscillations. The red curve shows the fit to a decaying cosine function plus an exponential decay.

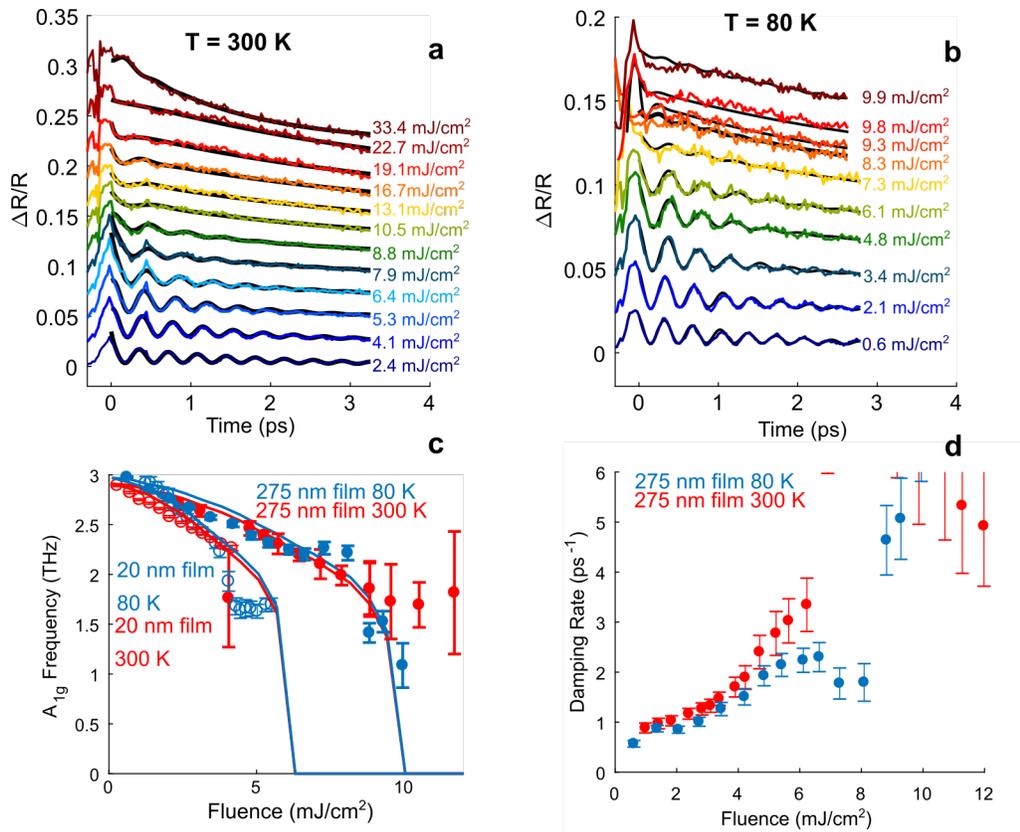

Fig. 2. **a** Single shot pump-probe traces for 275-nm Bi films with an initial temperature of 300 K. **b** Single shot pump-probe traces for 275-nm Bi films with an initial temperature of 80 K. **c** the coherent phonon frequency at 300 K and 80 K as a function of fluence for two separate films, one 20 nm thick and one 275 nm thick. Lines are simulation curves calculated using the photoexcited carrier density from an extended two-temperature model[27] and the carrier density-dependent anharmonic lattice potential[4]. At high fluences, the fitting breaks down when the coherent phonons vanish (above 11 mJ/cm$^2$ in the 275 nm film at 300 K and above 10 mJ/cm$^2$ at 80 K). The probe wavelength for all traces in this figure was 800 nm, and traces are offset for clarity. **d** The damping rate of the coherent oscillations, which increases with photoexcitation fluence. The fit starts to break down near 8 mJ/cm$^2$ incident fluence at 300 K, and near 9 mJ/cm$^2$ at 80 K.

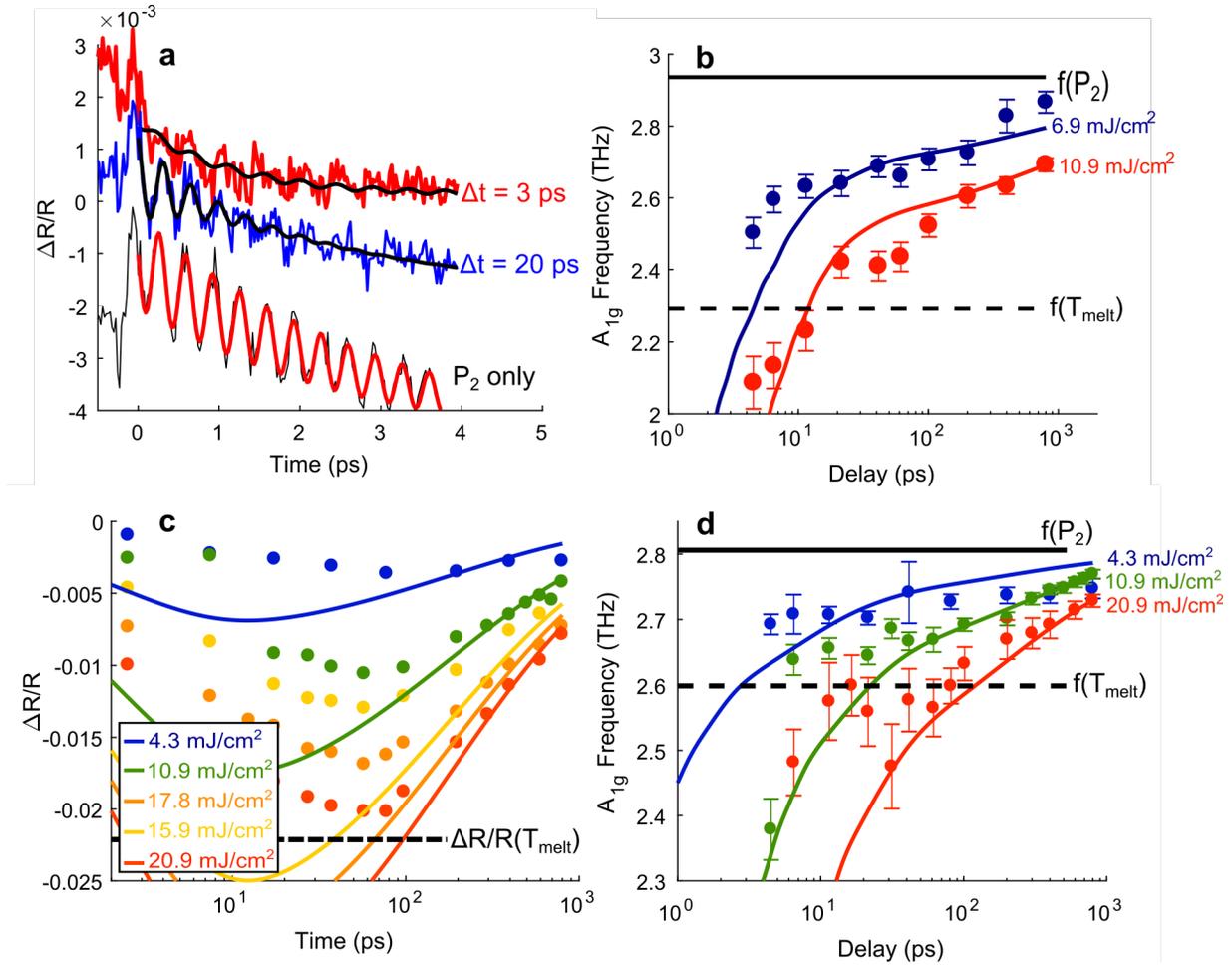

**Fig. 3** Two-pulse experiments on the highly excited state of the 275-nm bismuth film. **a** Single-shot traces following an initial excitation pulse ($P_1$) of 10.9 mJ/cm$^2$ and a second excitation pulse ($P_2$) of 1 mJ/cm$^2$ on a 275 nm film at 80 K. The coherent phonon signal is suppressed 3 ps after the first excitation pulse, and returns 20 ps later. $\Delta t$ is the time delay between the two excitation pulses. The probe wavelength for this experiment was 720 nm. **b** Coherent phonon frequency as a function of the time delay between an initial excitation pulse with incident fluence of 6.9 and 10.9 mJ/cm$^2$ and a second excitation pulse with incident fluence of $P_2$ =1 mJ/cm$^2$. The solid lines are the predicted oscillation frequency based on an extended two-temperature model. Within 1 ps after excitation at 10.9 mJ/cm$^2$, no phonon oscillations are induced by a weak second excitation pulse, indicating that the irradiated region of the film is in the symmetric phase. **c** The transient reflectivity of the Bi film at 300 K as a function of time after excitation for various pump fluences with an 800 nm probe. Note the logarithmic time scale. Solid points are measurements made at selected long times averaged over the 9 ps time window of the single-shot instrument. Solid lines are calculations based on an extended two-temperature model and the known thermoreflectance of bismuth. The simulations do not account for the reflectivity change due to hot carriers, and so are only expected to agree with experiment at long delay. **d** Transient $A_{1g}$ frequency measured using the same method as in panel **b** for the sample starting at 300 K for three representative pump fluences.

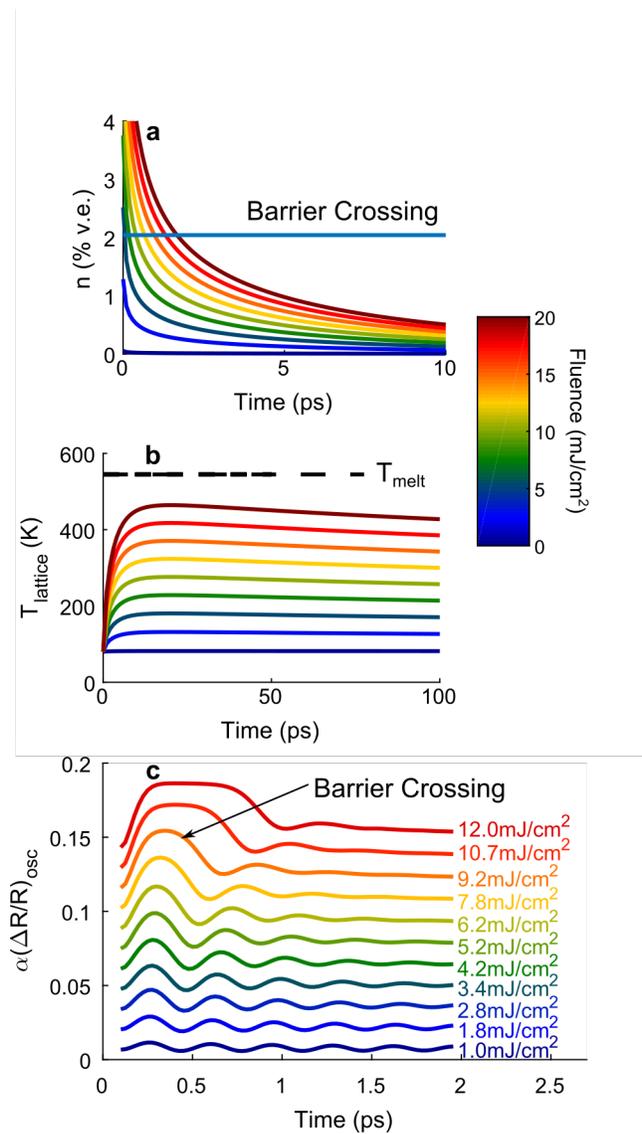

**Fig. 4** Extended two-temperature modeling in bismuth. **A** Shows the average carrier density within the optical penetration depth for the first 10 ps after excitation, for a range of fluences comparable to those used in experiment. The line in **a** corresponds to the carrier density required for bismuth atoms to be accelerated over the potential barrier at the center of the unit cell and enter the symmetric phase. **b** The lattice temperature in bulk Bi for the first 200 ps after excitation with an initial temperature of 80 K averaged over the penetration depth. The maximum lattice temperature is reached 15 ps after excitation. The line in a shows the melting temperature of bismuth (544 K). The melting temperature plus heat of fusion required to melt the surface is out of the range of the graph. **c** Simulations of the oscillatory part of the reflectivity obtained by solving the classical equations of motion for the $A_{1g}$ mode on a time-dependent potential defined by the carrier density in **b**. The simulated trace at 10 mJ/cm$^2$ crosses the barrier at the center of the unit cell, forming the symmetric phase.

## Materials and Methods

To determine the thickness of the thin films, we performed spectroscopic ellipsometry measurements. The ellipsometric parameters for two incidence angles, along with the fits, are shown below. The fits indicate that the film thickness is 20 nm. The 275 nm film is thicker than the optical penetration depth in the spectral range of the ellipsometer, so its thickness must be determined by profilometry.

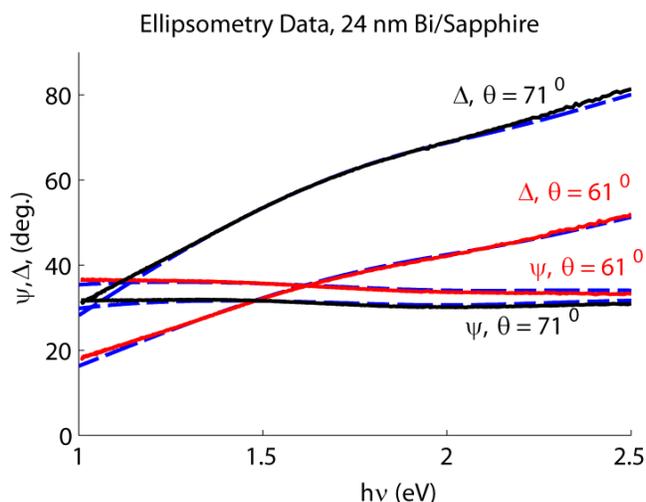

Figure S1: Ellipsometric Characteriztion of the 20 nm Bi film on sapphire. Measured spectorscopic ellipsometry parameters $\psi$ and $\Delta$, with experimental (blue) and fitted model (red, black) for incidence angles of $61^0$ and $71^0$.

To confirm the film thickness and roughness, we performed profilometry measurements on the films used. The films have a broad height distribution owing to their large crystallite size. However, the height distribution is well characterized. Profilometry Measurements confirm the height of the film. Thermal modeling uses the median height of the thicker film (275 nm).

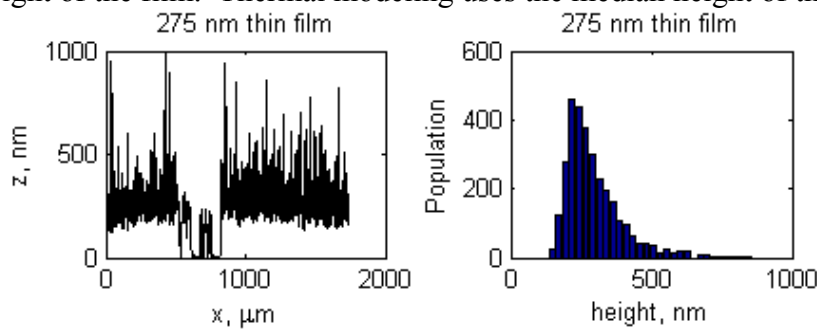

Figure S2: Profilometry Results for Bi thin films, 275 nm thin film. The polycrystalline films are rough, and have a broad height distribution. The effective height used in these calculations is the median height. Left: profilometry results along a 1.5 mm scan along the surface. Right: histogram of height distributions along the surface scan, showing film roughness.

Powder X-ray diffraction was used to confirm the crystallinity of the samples. PXRD measurements show the samples are polycrystalline, with no evidence of amorphous bismuth, or significant amounts of oxidation, and some texturing with the [111] (rhombohedral basis) plane preferentially ordered along the surface.

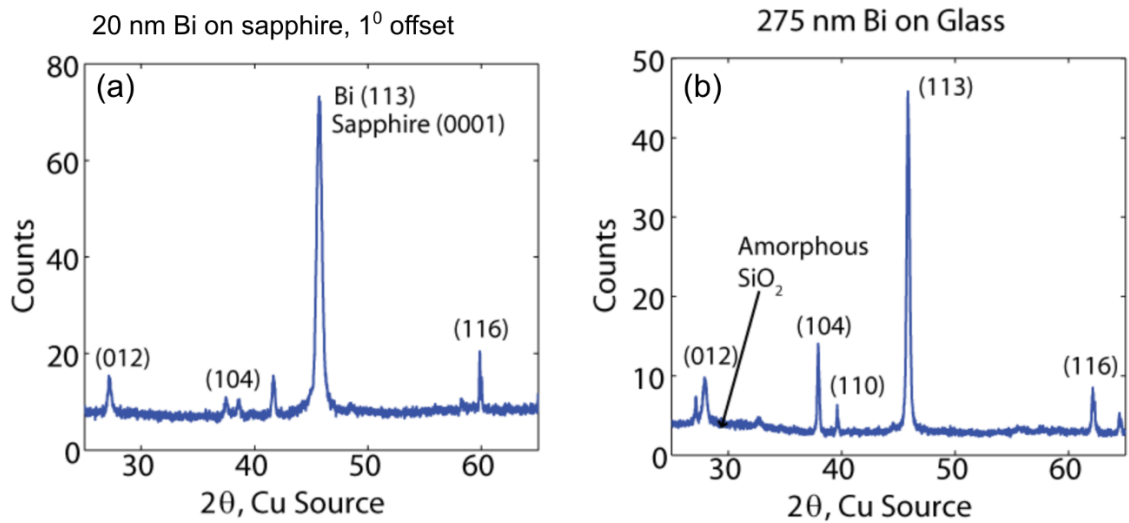

Figure S3: X-ray diffraction of Bi films. (a) The 20 nm Bi film on sapphire. (b) The diffraction pattern of the 275 nm thin film on glass. The films show texturing and are polycrystalline, the amorphous peak arises from the glass substrate. The strong peak at 45.7 deg. is from the sapphire substrate.

## Conventional Pump-Probe Measurements

Conventional pump-probe measurements were carried out with either an 800 nm probe, or with a white light continuum generated in sapphire. With the white light continuum, the spectrum was cut off with a 750 nm bandpass filter before the sample, and filtered with a 10 nm bandpass filter at the photodiode to select a specific wavelength. The pump pulse duration was the same as in single-shot measurements, approximately 70 fs. The probe was run at the half-harmonic of the laser repetition rate (500 Hz) and the probe was chopped to 250 Hz. The signal was detected on a lock-in amplifier (Stanford Research Systems SR810). The probe was chopped to determine an absolute value for $\Delta R/R$. Example traces of bismuth at 300 K are shown below.

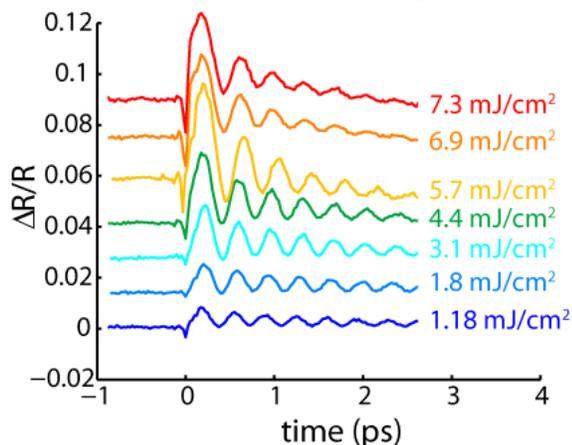

Figure S4: Conventional Pump-Probe Traces of the 275 nm bismuth film, with 800 nm pump and probe wavelengths. Fluences are labeled to the right of the trace, and traces are offset vertically for clarity.

## Dual-Echelon Single-Shot Spectroscopy

Single-shot measurements were carried out using a dual-echelon single-shot setup (Fig. S5) described in ref.[22]. The 800 nm fundamental of a 1 kHz Ti:sapphire amplified laser system (Coherent Legend) was downcounted to 10 Hz using a Pockels cell, and single pulses are selected using a shutter. For measurements where a Noncolinear optical parametric amplifier (NOPA) is used for the probe beam, 40% of the 800 nm beam from the Ti:sapphire laser system is directed into a NOPA tuned to 720 nm, and the remaining 60% is used as a pump beam. For measurements where an 800 nm beam is used as a probe, 10% of the beam is directed into the probe dual echelon system, and the remaining 90% is directed to the pump arm. The pump is horizontally polarized and directed onto the sample at approximately a 12 degree angle of incidence relative to the probe. The probe beam is vertically polarized, reflects off the sample surface at 45 degrees, and passes through a polarizer to reduce scattered light from the pump before the echelons are imaged onto a camera (Hamamatsu Orca-ER). An aperture in the Fourier plane of the sample further reduces pump scatter. Additional single-shot traces on the 20 nm film at 300 K, which is not presented in the main text of the paper, are shown in Fig. S6 below. The coherent oscillation frequency as a function of fluence of this data is shown in the main text, in Fig. 2c.

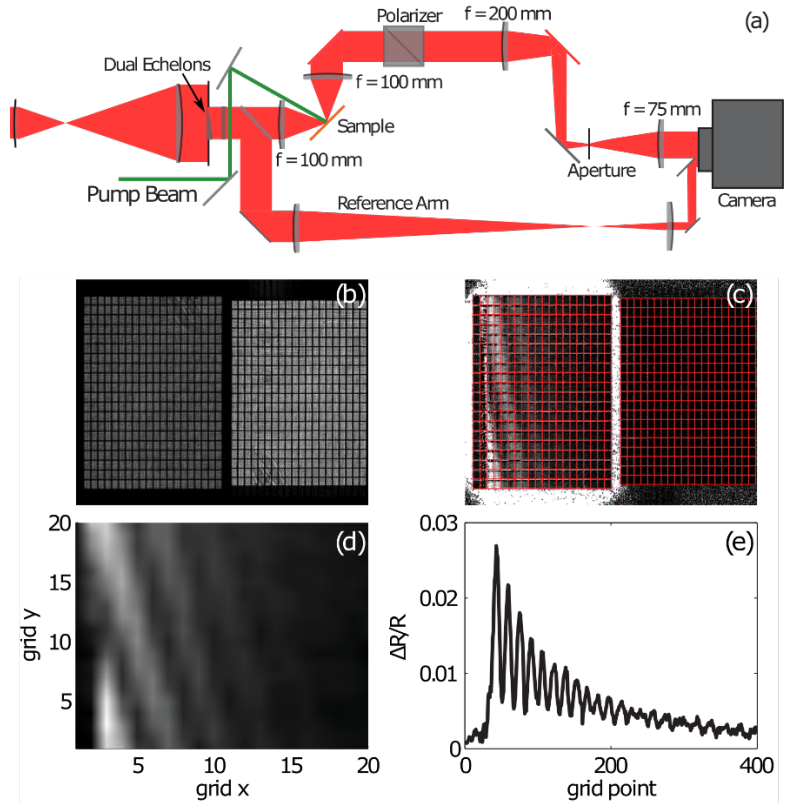

Figure S5: Dual-echelon single-shot spectroscopy. a Dual echelon spectroscopy setup. The beam transmitted through the echelons is focused onto the sample with a 100 mm fl lens, and then demagnified by approximately 2.5:1 onto the camera. A beamsplitter routes the echelon image around the sample, and produces a reference image of similar size on the camera. Two sets of images are taken, each with a signal and reference grid. One set (the signal) reflects off the irradiated sample, and a second set (the background) reflects off a sample with no pump irradiation. The images from the camera are saved to a computer, where the images are converted into a 400x1 array. b Raw background from the trace in Fig. 1. This image is the average over 100 laser shots. c The differential

image (the ratio of signal and background), along with the grids used for extracting the trace, shown in red. d The 20 by 20 grid of unfolded points, each averaged from the area within the grids shown in c. (e) The unfolded trace. Each step corresponds to 23.2 fs at 800 nm.

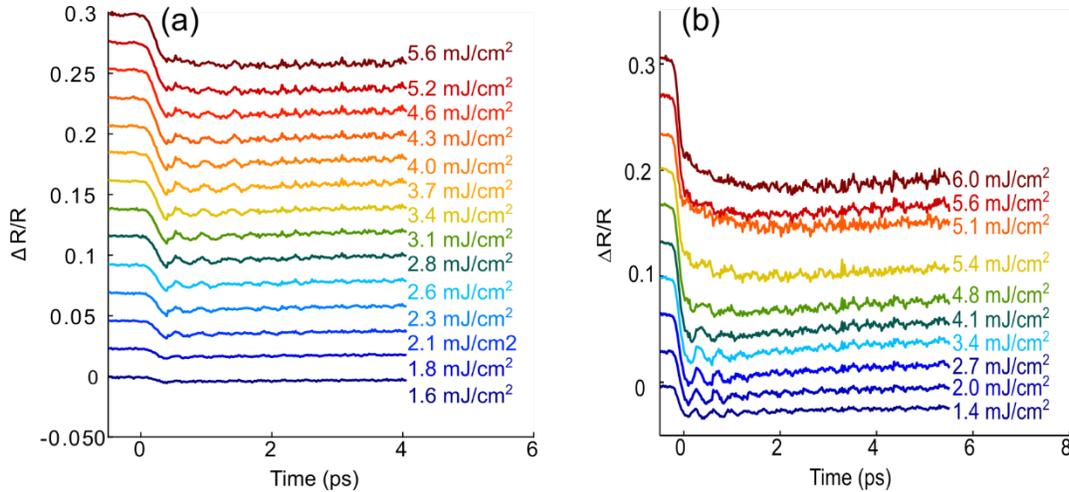

Figure S6. Single-shot reflectivity traces of the 20 nm film at room temperature (a) and 80 K (b). Coherent phonons vanish above 5 mJ/cm$^2$. Coherent phonon frequencies for these traces are shown in the main text in Fig. 2c.

## Laser Damage of Bismuth

When performing experiments above the damage threshold, a small number of laser shots can irradiate the sample without noticeable damage, even if the degree of pulsed illumination required to perform a pump-probe trace would damage the sample. In order to check for damage after a small number of laser shots, we imaged the sample surface after several measurements to ensure that there was no significant damage. The surface tends to darken before there is any significant change in the pump-probe trace, so we used the surface reflectivity as evidence for photodamage in bismuth. Images of the surface of a 275 nm thick film of bismuth before and after irradiation are shown in Fig. S7. Comparing the surface images before and after illumination with continuous exposure to 25 mJ/cm$^2$ pulses at 500 Hz repetition rate for several minutes to exposure to 50 pulses with a fluence of 25 mJ/cm$^2$, the surface damage due to exposure with 50 shots is considerably lower.

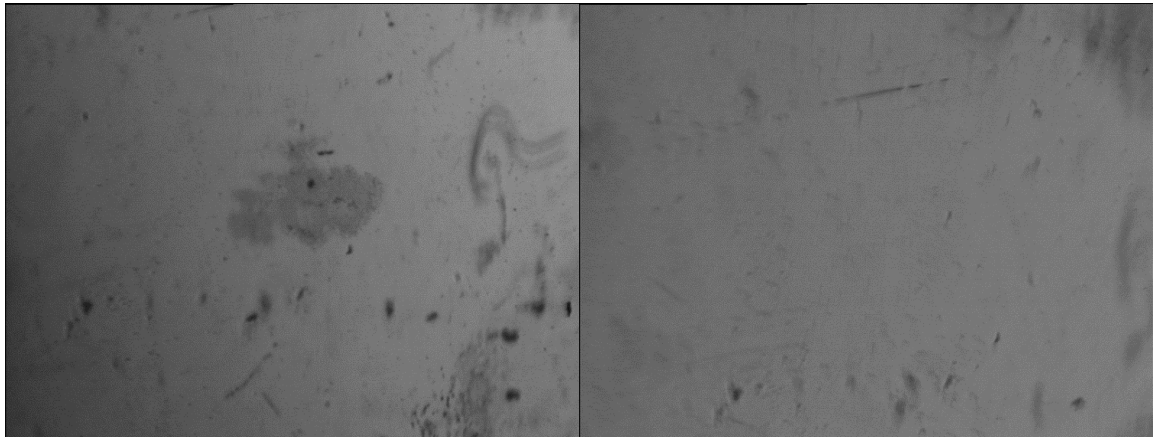

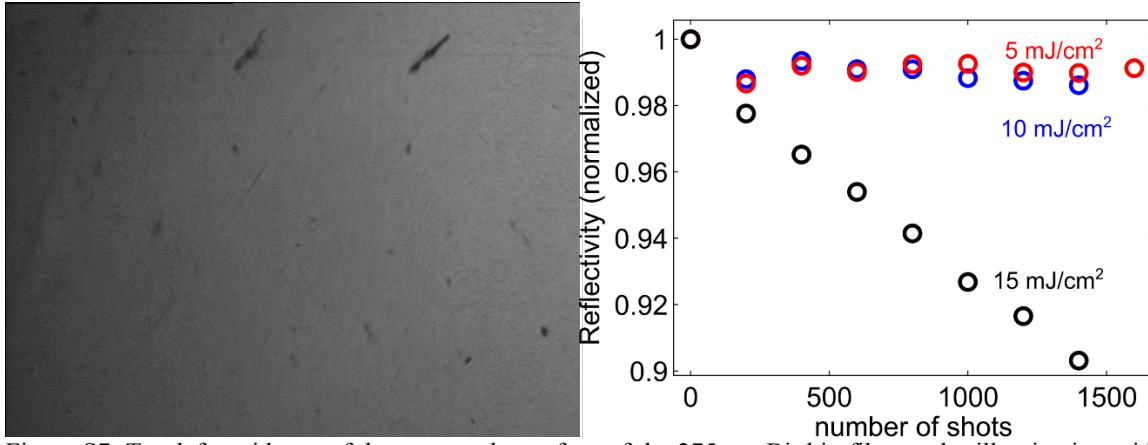

Figure S7: Top left: evidence of damage on the surface of the 275 nm Bi thin film, under illumination with a pump fluence of 25 mJ/cm² at 500 Hz. Top right: surface of 275 nm Bi film at 80 K, before irradiation. Bottom left: the sample surface after exposure to 50 shots at 20 mJ/cm² at room temperature. No damage is visible. Bottom right: reflectivity change as a function of the number of shots at 30 mJ/cm². Damage starts to appear after several hundred shots below 20 mJ/cm². All measurements in this work were taken with less than 50 shots per spot at 30 mJ/cm², and at less than 100 shots per spot at 20 mJ/cm².

## Two-temperature model

A modified two-temperature model was used to calculate the temperature rise in bismuth and the excited carrier density at the surface of the film. Our model is described in more detail in [27]. The equation for the carrier diffusion includes a carrier diffusion term, set by the carrier ambipolar diffusivity $D_e$, and a recombination time, $\tau$. We used a carrier recombination time of 15 ps [31] and an ambipolar diffusivity of 4,000 nm²/ps (40 cm²/s) which is in close agreement with literature values[31]. A set of three coupled differential equations describes the carrier density, $n_e$, carrier temperature, $T_e$, and lattice temperature, $T_l$.

$$\frac{\partial n_e}{\partial t} = D_e \frac{\partial^2 n_e}{\partial z^2} - \frac{n_e - n_0(T_l)}{\tau} \qquad 1.$$

$$C_e \frac{\partial T_e}{\partial t} = D_e \frac{\partial^2 T_e}{\partial z^2} - g(T_e)(T_e - T_l) \qquad 2.$$

The carrier thermalization term includes a carrier-temperature dependent electron-phonon coupling constant, $g(T_e)$. We assumed the electron-phonon coupling constant is independent of carrier temperature, in contrast to the model presented in [28]. We note that unlike in [28], our model includes both carrier density and carrier temperature as independent parameters, which is equivalent to assuming electrons and holes may have different chemical potentials. In addition, we use a different heat capacity for the excited electrons, which necessitates a different scaling of the electron-phonon coupling constant with carrier temperature to produce the same heating rate at constant incident fluence. Fig. S10 shows a comparison of our model and the model in Arnaud et al[28] for the heating of a 30 nm bismuth film with an excitation fluence of 1.11 mJ/cm². They are in good agreement, demonstrating consistent heating rates between the two models. As long as the carrier-temperature or carrier-density dependent bond softening is treated appropriately, both models can reproduce both the lattice heating rate and the photoinduced phase transition, and our experiment is essentially agnostic to the differences in these models.

$$C_l \frac{\partial T_l}{\partial t} = D_l \frac{\partial^2 T_l}{\partial z^2} + g_0 n_e (T_e - T_l) + \frac{E_g (n_e - n_0)}{\tau}$$

3.

The lattice thermalization term includes lattice diffusivity, temperature gain from electron-phonon coupling, and carrier recombination, where $C_l$ is the specific heat of the lattice, $D_l$ is the thermal diffusivity of the lattice, and $E_g$ is the band gap. The initial carrier concentration for a uniformly distributed film can be described as

$$n_e(z, t = 0) = \frac{F(1-R)}{E_{ph} d n_{VE}} + n_0(T_0)$$

4.

Where $d$ is the depth of the film (25 or 275 nm), $R$ is the reflectivity of bismuth at 800 nm (0.7), $E_{ph}$ is the energy of an incident photon (1.55 eV), $n_{VE}$ is the initial concentration of valence electrons, and $n_0$ is the thermal population of electrons in the conduction band (0.1% at 298 K). The result is a carrier concentration in units of fraction of valence electrons excited. In general for our experiments, $n_e(t=0) \gg n_0(T_0)$. The initial carrier temperature can be approximated as

$$T_e(z, t = 0) = \frac{E_{ph} - E_g}{C_e k_b} \times \frac{n_e}{n_e + n_0(T_0)}$$

5.

Where $k_b$ is Boltzmann's constant, $E_{ph}$ is the photon energy, $E_g$ is the indirect band gap (0.2 eV) and $C_e$ is the electronic heat capacity. This results in the correct overall temperature rise of the system.

The boundary conditions for the two-temperature model control the surface and interfacial conductivity. We assume that the thermal conductivity of air is sufficiently small to be neglected on our timescales, and that the substrate remains close to room temperature.

$$C_l \frac{\partial T_l(z_{end})}{\partial t} = C_{TI}(T_l - T_0)$$

6.

The parameters used in our model are summarized below. They agree roughly with the results of [28,31,32].

| Paramater | Symbol | Value |
|---|---|---|
| Band Gap | $E_g$ | 0.2 eV |
| Photon Energy | $E_{ph}$ | 1.55 eV |
| Equilibrium Carrier Density | $n_0$ | 0.001 % |
| Electron Specific Heat | $C_e$ | 2.07 $10^{-23}$ J/K/e$^-$ |
| Lattice Specific Heat | $C_l$ | 1.18 $10^{-21}$ J/K/nm$^3$ |
| Carrier Recombination Time | $\tau$ | 15 ps |
| Electron-Phonon Coupling Constant | $g_0$ | 3x$10^{15}$ W m$^{-3}$ K$^{-1}$ [27,28] |
| Thermal Interface Conductivity | $C_{TI}$ | 1,950 W/cm$^2$/K [23] |

| Carrier Diffusion Constant | $D_e$ | 1,300 nm$^2$/ps [31] |

Numerical solutions for the 20 nm and 275 nm films as a function of fluence for short and long timescales are shown below. Starting at 80 K, it takes 20 mJ/cm$^2$ for the 275 nm film to reach the melting point under this model, and over 50 mJ/cm$^2$ to reach the melting point plus the heat of fusion.

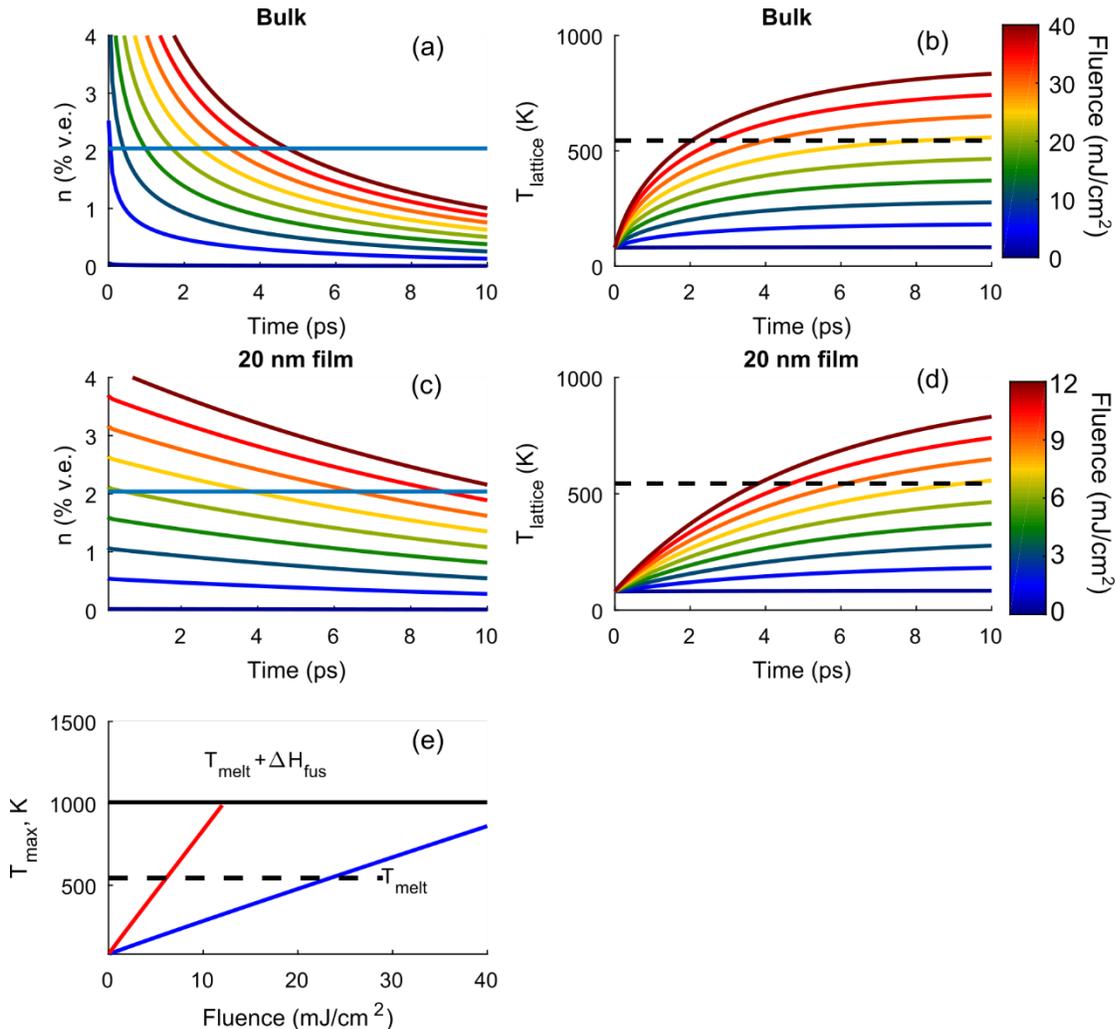

Figure S8: (a): Two temperature model results for the 275 nm film, showing the carrier density at short times (a) and lattice temperature near the surface at short times (b) The initial temperature of the films is 80 K. Results for the carrier density and lattice temperature of the 20 nm film are shown in (c) and (d). The maximum lattice temperature as a function of fluence for both the bulk (275 nm film) and the 20 nm is shown in (e). The lattice temperature reaches the melting point at comparable fluences to formation of the symmetric phase, but does not surpass the heat of fusion before onset of the symmetric phase.

Figure S9(a) shows a plot of the predicted phonon frequency *vs.* carrier density from DFT calculations performed by Murray et al (left), and Fig. S9(b) the predicted phonon frequency over the first 50 ps (log time scale, right) calculated using equation S7. In Fig. S9(b), the phonon frequencies at long time delays are offset at long times due to a temperature rise[25]. The redshift of the $A_{1g}$ phonon frequency is calculated as a function of carrier density and lattice temperature

$$F(n_e, T) = F(n_e) - 8.4 \times 10^{-4}(T - 300\ K) \qquad 7.$$

Where $F(n_e)$ is the phonon frequency predicted by DFT calculations at room temperature. This assumes that the bond softening due to lattice temperature is independent of the bond softening due to carrier density to first order.

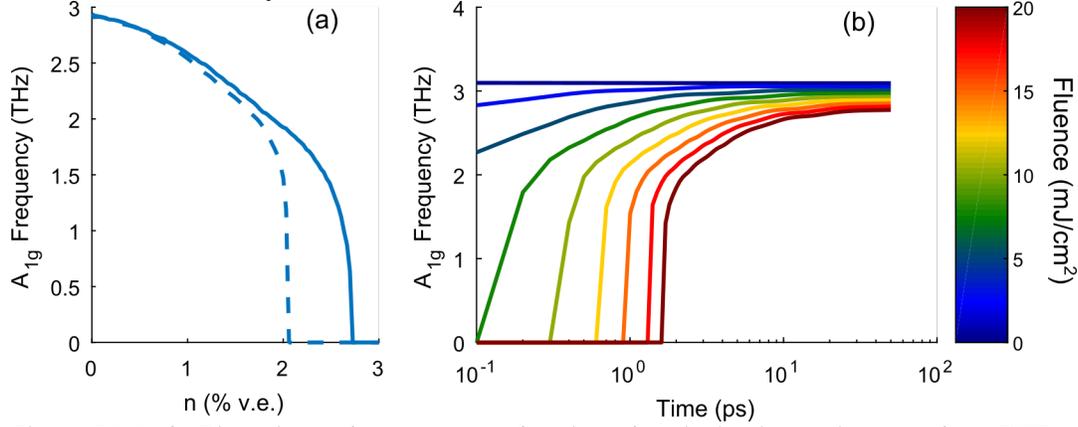

Figure S9: Left: Bismuth $A_{1g}$ frequency as a function of excited valence electrons, from DFT calculations. The dashed line represents the harmonic frequency (the curvature of the lattice potential about the local minimum) and the solid line is the anharmonic frequency (the inverse of the period of motion for an impulsively excited sample from the ground state). The anharmonic frequency drops to zero when the bismuth crosses the center of the unit cell. Above this point, the mode ceases to be Raman active to first order. The harmonic frequency drops to zero when the curvature of the lattice potential at the local minimum goes to zero, and the lattice distortion vanishes. Right: Phonon frequency as a function of time for the 275 nm film after impulsive excitation for the same set of pump fluences used in Fig. S7, left, calculated using eq. S7. Barrier crossing occurs when the calculated phonon frequency drops to zero THz for more than 150 fs (above 10 mJ/cm$^2$).

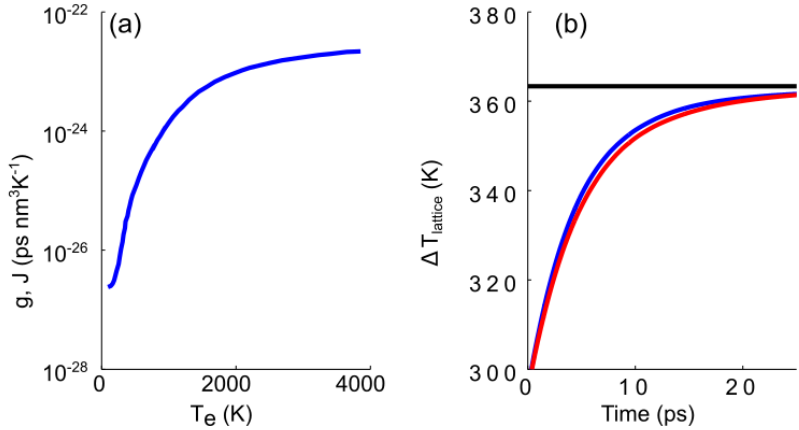

Figure S10: (a) Carrier temperature dependent electron-phonon coupling constant. (b) comparison between a conventional[28] and extended two-temperature model under the same experimental conditions. The red line shows the conventional two-temperature model, where carrier entropy is fixed and the electron and hole chemical potentials are equal. The blue curve shows the extended two-temperature model used in this work. Both curves were calculated for an 800 nm pulse with a fluence of 1.11 mJ/cm$^2$ incident on a 30 nm thin film. The black line shows the final temperature value based on the film thickness and deposited pulse energy, assuming uniform heating. The final temperature and rise time agree to within 10% % between the two models.